\documentclass[aip,apl,reprint]{revtex4-1}

\usepackage{graphicx}
\usepackage{amsmath}
\usepackage[version=4]{mhchem}
\usepackage{url}
\usepackage{caption}
\usepackage{siunitx}
\begin{document}

\title{Thickness-dependent secondary-electron emission from suspended \ce{MoS2} membranes in the helium ion microscope}

\author{Cyan Kim}
\author{David Lister}
\author{Philip Jackle}
\author{Karen L. Kavanagh}
\email{kavanagh@sfu.ca}
\affiliation{Department of Physics, Simon Fraser University, Burnaby, BC, Canada}

\date{\today}

\begin{abstract}
Secondary-electron (SE) emission in the helium ion microscope (HIM) becomes sensitive to membrane thickness when the sample is thin enough for He-ion transmission and when the SEs emitted from the bottom surface are collected. We correlated the total SE intensity of suspended, nanometer-thick $\mathrm{MoS}_2$ flakes on lacey carbon with thickness measured independently by electron energy-loss spectroscopy. The response peaks at 40--55~nm, with an apparent back-to-front SE signal ratio reaching 4.7. The peaked, thickness-dependent component is attributed primarily to SE emission at the bottom surface of the flake, rather than to transmitted ions striking instrument surfaces. Applying SRIM ionization profiles, an asymmetric SE-escape model with a longer escape depth on the exit side reproduces the response. We find an effective exit-side escape depth of $\approx$~10~nm, five times the assumed 2~nm entrance value, suggesting that deposited energy reaches the exit surface far more efficiently than the entrance surface or that the SRIM model's energy deposition profile is shifted by an effect such as channeling. The correlation provides a rapid thickness screen for suspended membranes and a route to testing low-energy ion--solid interaction models in thin materials.
\end{abstract}

\maketitle
Scanning electron microscopy (SEM) and He ion microscopy (HIM) both rely on the emission and detection of low-energy secondary electrons (SEs) for imaging. This is usually limited to SE emission from the top surface of a thick sample, emitted backward relative to the beam. In this mode, the technique is highly surface sensitive and can detect monolayer thickness variations such as the number of graphene layers on a support.\cite{Hiura2010}

SE emission will also occur at the bottom or exit side if the sample is a self-supporting film thin enough for the ions to transmit.\cite{Oppel1972, Meckbach1975} For ion beams sent through such films, the backward and forward SE yields, $\gamma_\mathrm{B}$ and $\gamma_\mathrm{F}$, respectively, have been measured and modeled for many decades.\cite{Meckbach1975, Rothard1989, Zhurenko2019, Pauly2003} For carbon foils traversed by protons (25--250~keV), $\gamma_\mathrm{F}$ exceeds $\gamma_\mathrm{B}$: the ratio increases roughly linearly with beam energy, from $\approx$~1.2 at the lowest energies to 1.55 above 140~keV.\cite{Meckbach1975} Later studies of energetic molecular and heavy ions, including the dependence on foil thickness, found forward-to-backward ratios that stayed of order 1--2 throughout the accessible range.\cite{Rothard1990mol, Rothard1990, Billebaud1995, Rothard1995} In all of these experiments the foil was much thinner than the ion range, so only the rising portion of the thickness dependence---before any maximum---was accessible. Within kinetic-emission theory, in which the SE yield scales with the electronic energy deposited near each surface,\cite{Sternglass1957} this forward excess is attributed primarily to energetic recoil ($\delta$) electrons---target electrons struck hard enough to travel forward---transporting energy toward the exit surface.\cite{Rothard1992}

These earlier experiments used broad, unfocused beams, recording yields from one nominally uniform foil at a time, without imaging. Additionally, they involved much higher projectile energies per nucleon than typical of HIM.\cite{Rothard1990} In the HIM, transmission configurations have characterized ion scattering through free-standing membranes,\cite{Marshall2012, Fox2013, Hall2013, Woehl2016, Wang2018, Emmrich2021} and measured thickness-dependent SE emission from supported \ce{Si3N4} membranes.\cite{Petrov2018} The present experiment images suspended \ce{MoS2} membranes with independently measured thickness and separates the top-surface and bottom-surface SE contributions, extending front- and back-emission studies to the low-energy HIM regime.

Suspended \ce{MoS2} membranes tens of nanometers thick are suitable targets: at 30~keV this thickness range spans the transition from nearly full He transmission to complete stopping. Flakes in this range serve as active layers in membrane-based diode sensors, where focused helium beams provide patterning and tuning.\cite{Fox2015} In our previous sensor work,\cite{Fawzy2024} selecting flakes of suitable thickness required separate atomic force or transmission-electron microscopy. A faster thickness assessment in the imaging instrument would remove that step.

In this Letter, we report the collected SE signal of suspended \ce{MoS2} flakes (10--200~nm thick), measured from focused-ion-beam images, as a function of the local thickness, $t$. The flakes, supported on a lacey-carbon grid, were imaged with a 30~keV He$^+$ beam in a transmission geometry,\cite{Kavanagh2017} with $t$ determined independently by electron energy-loss spectroscopy (EELS).

For these measurements, flakes were prepared from natural \ce{MoS2} crystals mined in Norway: a bulk fragment was sonicated in isopropyl alcohol to induce exfoliation, and after centrifugation the supernatant was drop-cast onto a lacey-carbon copper TEM grid (3.05~mm diameter) and dried under ambient conditions.

 \begin{figure}[htbp]
  \centering
  \includegraphics[width=0.92\linewidth]{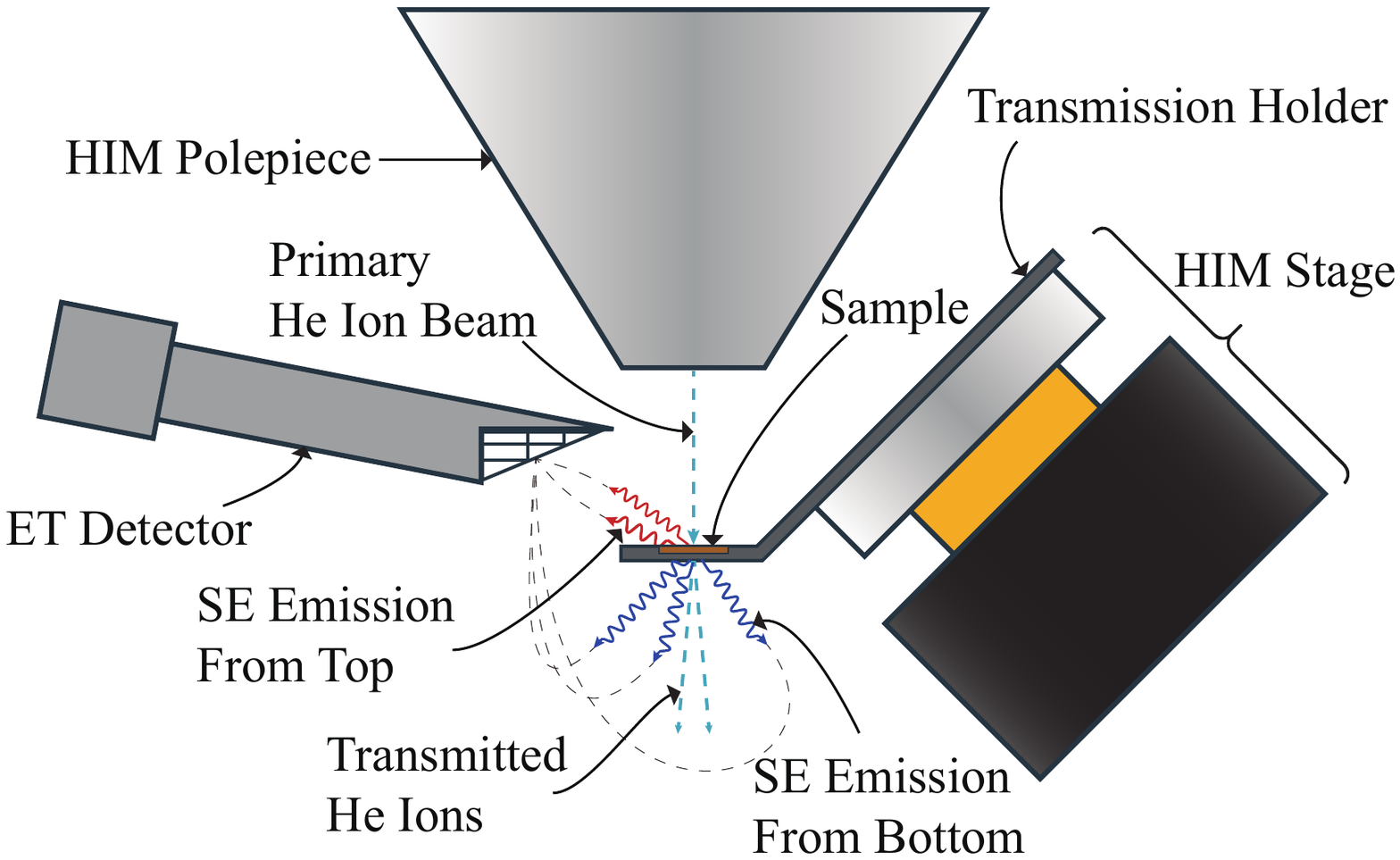}
  \caption{Schematic diagram of the transmission HIM geometry. The stage is tilted to open a downstream path, while the TEM grid at the holder tip remains perpendicular to the primary beam (normal beam--membrane incidence).}
  \label{fig:geometry}
\end{figure}

Measurements were performed using a HIM (Zeiss Orion NanoFab) fitted with a modified transmission sample holder on the conventional tilt stage, as shown in Fig.~\ref{fig:geometry}.\cite{Kavanagh2017} The stage is tilted to open a downstream path, while the holder keeps the beam at normal incidence to the membrane. SEs emitted from both the top and bottom surfaces are collected, while transmitted ions pass through the open holder.

The primary helium ion beam was operated at an accelerating voltage of 30~kV with a GFIS pressure of $1.00 \times 10^{-7}$~Torr and a 20~\textmu m molybdenum aperture, yielding a beam current of $0.13 \pm 0.02$~pA measured with the in-chamber Faraday cup. These conditions corresponded to per-image fluences of $8 \times 10^{13}$ (30~\textmu m field of view) to $9 \times 10^{14}$~He$^+$/cm$^2$ (8.5~\textmu m), below the $\sim 10^{16}$~He$^+$/cm$^2$ onset of detectable lattice damage in \ce{MoS2} and far below the $\sim 2.6 \times 10^{18}$~He$^+$/cm$^2$ complete-milling dose.\cite{Fox2015} These are low-fluence, though not damage-free, imaging conditions (Sec.~SI).

Critically, detector gain and black-level settings were held fixed for all images, and all analyzed pixels lay within the unclipped 8-bit range, so grayscale served as a relative measure of detector signal. The decomposition and modeling below assume an approximately linear detector response with the black level at zero, controlled but not independently calibrated (Sec.~SI of the supplementary material). SE detection used a standard Everhart--Thornley (ET) photomultiplier-based detector located $\sim$16$^{\circ}$ above the horizontal stage plane, and $\sim$9~mm from the sample to the center of its Faraday cage. Images were acquired in line-scan mode, $2048 \times 2048$ pixels, 100~\textmu s dwell, 20~\textmu s line retrace, two-frame averaging, over fields of view of 8.5--30~\textmu m.

Reference flake thicknesses, $t$, were determined by EELS (Gatan Enfinium spectrometer) within a field emission, scanning TEM (FEI Osiris) at 200~kV, using the log-ratio relation $t = \lambda \ln(I_\mathrm{t}/I_0)$, with $I_\mathrm{t}$ the total spectral intensity, $I_0$ that of the zero-loss peak, and $\lambda \approx 132$~nm the inelastic mean free path from the Malis parameterization for these conditions.\cite{Malis1988} Comparison with the Iakoubovskii parameterization assigns a systematic uncertainty of the order of 15--20\% to the thickness scale, in addition to $\sim$10\% point-to-point repeatability (Sec.~SIII).\cite{Iakoubovskii2008} This systematic uncertainty rescales all thicknesses coherently, shifting the maximum without altering the curve shape.

\begin{figure}
    \centering
    \includegraphics[width=1\linewidth]{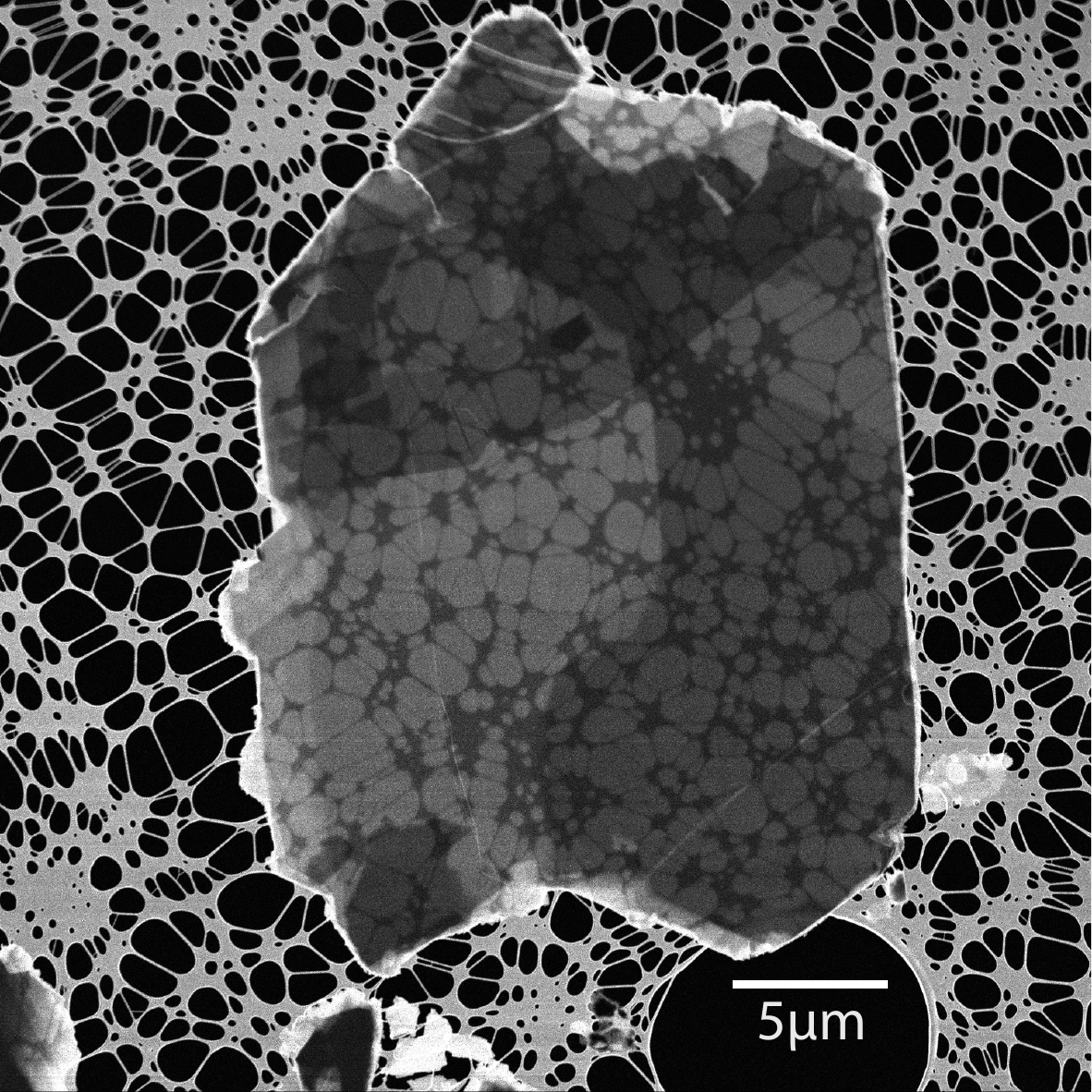}
    \caption{HIM SE image of a suspended \ce{MoS2} flake on the lacey-carbon support, acquired in the transmission geometry.}
    \label{fig:HIMimage}
\end{figure}

Fig.~\ref{fig:HIMimage} shows an example SE image of a \ce{MoS2} flake. Quantitative intensities came from polygonal regions of visually uniform contrast on the flake and bare lacey-carbon support within the same image. Region-of-interest (ROI) sizes varied with the available uniform region; each ROI mean entered the analysis as one observation (Secs.~SIV and SV). The \ce{MoS2} mean signals were normalized to the carbon reference of the same image, giving the normalized SE intensity, $I_\mathrm{n} = \bar{I}_{\mathrm{MoS_2}}/\bar{I}_\mathrm{C}$. Normalizing within each image removes image-to-image variations in beam current and detector response; because the carbon support itself transmits ions, however, self-normalization also cancels any response common to flake and support, and cross-configuration comparisons below must be read with that partial blindness in mind (Sec.~SII).

\begin{figure}
    \centering
    \includegraphics[width=8.5cm]{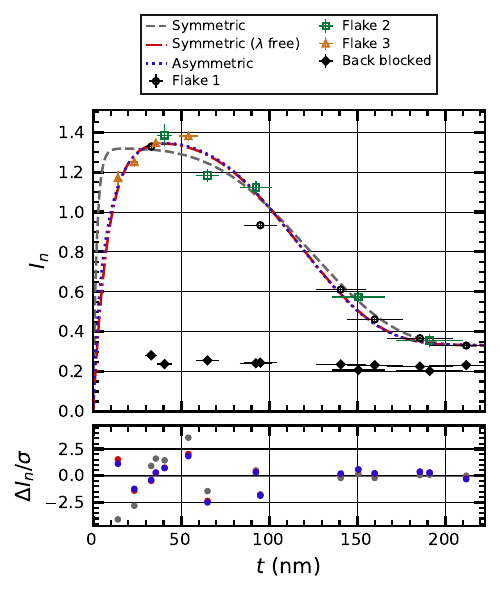}
    \caption{Symmetric (fixed \qty{2}{\nano\meter} escape depth), symmetric free-$\lambda$, and asymmetric SE-escape models compared with the pooled HIM--EELS data (vertical bars, 68\% credible intervals; horizontal, 10\% point-to-point thickness repeatability); black diamonds show the backside-blocked control. Lower panel, normalized residuals under common weighting. The free-$\lambda$ and asymmetric curves nearly coincide: the shape of the data does not choose between them, and the free-$\lambda$ fit is excluded on physical grounds alone (Sec.~SVIII).}
    \label{fig:model-comparison}
\end{figure}

Fig.~\ref{fig:model-comparison} shows the normalized intensity $I_\mathrm{n}(t)$, pooled from the three analyzed flakes. The intensity increases with thickness to a maximum, $I_\mathrm{n} \approx 1.35$ near 40--55~nm, then decreases to a low, nearly constant level for the thickest regions.

The collected signal can contain four contributions: top-surface emission from the primary beam, $\mathrm{SE1_{top}}$;\cite{Ramachandra2009, Bell2009} membrane bottom-surface emission, $\mathrm{SE1_{bottom}}$, when transmitted ions deposit energy near the bottom surface;\cite{Oppel1972, Meckbach1975, Petrov2018} and instrument emission from transmitted ions striking surfaces below the sample, either from the direct, unscattered beam, $\mathrm{SE3_{beam}}$, or from ions scattered in the flake or support, $\mathrm{SE3_{scattered}}$.\cite{Hall2013, Beyer2015} To determine which of these four carries the thickness-dependent signal, we blocked the exit side to isolate the top-surface channel, and added a copper-tape converter to test the scattered-ion channel.

Two channels can be identified directly. At large thickness, no ions transmit, and the signal is only $\mathrm{SE1_{top}}$. At zero thickness---the support holes---the only possible source is $\mathrm{SE3_{beam}}$, yet the holes appear black in every transmission image. Therefore, $\mathrm{SE3_{beam}}$ is undetected. This is consistent with the unscattered beam entering the lower pumping port ($\sim$14.2$^{\circ}$ semi-angle) and traveling more than 30~cm before meeting any surface (Sec.~SVII). When there is transmission, the on-flake signal is $\mathrm{SE1_{top}}$ + $\mathrm{SE1_{bottom}}$ + $\mathrm{SE3_{scattered}}$.

To isolate $\mathrm{SE1_{top}}$, the same regions of two of the three flakes were re-imaged with the backside of the grid blocked by copper tape, applied with its adhesive (non-metallic) side facing the sample to further suppress tape SE generation. Because the support itself is also blocked, each blocked dataset was normalized to the carbon reference of its paired transmission image (Sec.~SVI). Across the 11 re-imaged thickness groups, the blocked intensities are weak and nearly thickness-independent, with mean $I_\mathrm{n}=0.24$ and standard deviation 0.02. Since blocking leaves only $\mathrm{SE1_{top}}$, the remainder of the transmission-geometry signal---the exit-channel signal---must be generated by the transmitted ions.

Blocking the backside removes $\mathrm{SE1_{bottom}}$ and $\mathrm{SE3_{scattered}}$ together, so it cannot determine whether $\mathrm{SE1_{bottom}}$ or $\mathrm{SE3_{scattered}}$ dominates the exit-channel signal. To separate them, we enhanced only the scattered channel: if $\mathrm{SE3_{scattered}}$ contributes strongly, deliberately amplifying it should visibly increase the signal.

A copper-tape strip was attached vertically below the stage adjacent to the sample aperture, with its bare copper face toward the ET detector; scattered transmitted ions strike it at grazing incidence (4--7$^{\circ}$ relative to the tape plane), while the unscattered beam passes parallel into the pumping aperture (Sec.~SII, Fig.~S1). From this geometry and the SRIM exit-angle distributions, the strip intercepts $\approx$30\% of the transmitted ions at the peak thicknesses, and doubling its area raises the intercepted fraction by only $\approx$~1.1$\times$ (Sec.~SII).

Measurements with no tape, a small tape, and a large (doubled-area) tape changed the on-flake normalized signal only modestly: by $+5\% \pm 5\%$ and $+15\% \pm 10\%$, respectively (mean $\pm$ standard deviation across the 15 matched thickness points), leaving the peak position unchanged (Fig.~S2). Had $\mathrm{SE3_{scattered}}$ dominated the exit-channel signal, with intercepted ions generating detected SEs at least as efficiently as those striking remote chamber surfaces, the small tape alone would have raised the signal by $\gtrsim$25\%; and the tripled response on doubling the area, far exceeding the $\approx$1.1$\times$ increase in interception, instead points to the acquisition-order drift (Sec.~SII). By contrast, a tape tilted to 45$^{\circ}$ into the direct beam produced a bright-field image in which the support holes---black in every regular image---showed strong signal well above the black level (Fig.~S3): a converter in the beam path makes $\mathrm{SE3_{beam}}$ readily detectable, a positive control showing that SEs generated below the stage are collected.

Quantitatively, the exit-channel signal at the peak is 4.7 times the blocked level---the apparent back-to-front signal ratio---and the total peak signal is 5.7 times the blocked level. Both are cross-acquisition ratios subject to uncorrected beam-current and detector drift, and both are collected-signal ratios, not intrinsic yields: blocking suppresses all three non-top channels together (Sec.~SVI). $\mathrm{SE1_{bottom}}$ and $\mathrm{SE3_{scattered}}$ together supply most of the signal between roughly 15 and 150~nm. Consistently, the thickest transmission-geometry regions ($t = 186$--212~nm, transmitted fraction $\approx$3--8\%, Sec.~SVII) level off at $I_\mathrm{n} = 0.26$--0.29, within $\approx$0.05 of the blocked level---a within-acquisition check that bounds drift and the residual transmitted-ion signal together.

The identification now follows by elimination. The blocked control isolates $\mathrm{SE1_{top}}$ and shows it is small and thickness-independent. The black support holes show that $\mathrm{SE3_{beam}}$ is undetected; the exit-channel signal must therefore come only from $\mathrm{SE1_{bottom}}$ or $\mathrm{SE3_{scattered}}$. If $\mathrm{SE3_{scattered}}$ were the dominant source, the converter should have amplified the signal strongly; it did not. What remains is $\mathrm{SE1_{bottom}}$: the bottom surface of the flake carries the thickness-dependent SE signal (Sec.~SII).

In previous fast-ion foil experiments, the forward excess has been attributed to $\delta$ electrons ejected with enough energy to reach and cross the exit surface;\cite{Rothard1992} if that mechanism operated here, it would explain a bottom-surface enhancement. At HIM energies, however, kinematics rules it out. The maximum energy a projectile of mass $M$ and energy $E$ can transfer to a stationary quasi-free electron of mass $m_e \ll M$ is $T_\mathrm{max} = 4(m_e/M)\,E = 2m_ev^2 \approx 16$~eV for 30~keV He; even for a valence electron already moving at a Fermi velocity ($E_F = 5$--10~eV), a head-on collision transfers at most $\approx$40~eV, compared with hundreds of eV in the fast-ion experiments. Energetic $\delta$ rays are therefore absent. Low-energy excitations---slow SEs, electron--hole pairs, and plasmons---are still generated in abundance, and how far they carry energy toward the exit surface is the question the following model quantifies.\cite{Bell2009, Hlawacek2014}

To model this, we used SRIM\cite{Ziegler2010} to simulate 30~keV He transport through \ce{MoS2} and took the resulting electron ionization data as input for predicting the change of SE emission with thickness. We compared three models: a symmetric SE-escape model with the escape depth fixed at 2~nm, which serves as the null hypothesis; the same symmetric model with the escape depth left free; and an asymmetric model in which the exit side has its own escape depth. Common to all three are an overall horizontal stretch, $s$, which accommodates the systematic depth-scale difference between EELS and SRIM (Secs.~SIII and SVII), and an exponentially weighted SE extraction: a decaying exponential applied from either the top or the bottom surface weights the ionization profile to infer the SE emission. The top and bottom escape functions are parameterized as
\begin{align}
    Y_t(t; \lambda, s) &= \int_0^t D(z/s)\,e^{-z/\lambda}\,dz,
\label{eq:kernels}\\
    Y_b(t; \lambda, s) &= \int_0^t D(z/s)\,e^{-(t-z)/\lambda}\,dz,\nonumber
\end{align}
where $t$ is the measured thickness, $\lambda$ an SE escape depth in measured nanometers, $s$ the stretch, and $D$ the electron-ionization energy density from SRIM, computed for a 300~nm target with $10^{6}$ ions for minimal Monte Carlo noise (a recoil-corrected deposition changed the results negligibly; Sec.~SVIII).

The null hypothesis is that the top and bottom of the flake share one escape depth, fixed at the physical value $\lambda_{esc} =
\qty{2}{\nano\meter}$,\cite{Ramachandra2009, Chee2016, Saitoh2025} with the
only added asymmetry a less efficient SE collection from the back:
\begin{equation}
    I_\mathrm{n}^{\mathrm{sym}}(t) = C_o \left [ Y_t(t;\lambda_{esc},s)  + C_{b}\,Y_b(t; \lambda_{esc},s)\right ],
\label{eq:sym}
\end{equation}
where $C_o$ is the overall vertical scale and $C_b$ the back collection penalty. This model should fit if the SRIM profile shape is accurate and the deposited energy thermalizes near the deposition site. The free-$\lambda$ variant is Eq.~(\ref{eq:sym}) with the shared depth fitted rather than fixed.

The asymmetric model considers the effect if the back side has a different effective escape depth from the front,
\begin{equation}
    I_\mathrm{n}^{\mathrm{asym}}(t) = C_o \left [ Y_t(t;\lambda_{esc},s)  + C_{b}\,Y_b(t; \lambda_{back},s)\right ],
\label{eq:asym}
\end{equation}
with $\lambda_{back}$ free. Physically, a longer back-side depth could result from energy carried forward with momentum by recoil electrons, or $\lambda_{back}$ can act as an effective correction to the deposition profile for effects SRIM does not consider, such as channeling or the material's anisotropy. The entrance scale is fixed to avoid over-defining this model (freeing both scales and $C_b$ together leaves the fit unable to separate them) and is retained in the symmetric model for consistency; the fitted exit-side depth and stretch are insensitive to varying it over 1.5--2.5~nm, while $C_b$ shifts with it (Sec.~SVIII).

Comparing the models in Fig.~\ref{fig:model-comparison}, the symmetric model does not describe the data ($\chi^2/\nu=48.2/12=4.02$), while the asymmetric model fits better ($\chi^2/\nu=14.9/11=1.35$) with a fitted exit-side escape depth of \qty{9.6(1.3)}{\nano\meter}. Interpreted as recoil-electron transport, this depth would require electron energies far above the $\approx$40~eV kinematic ceiling, or a significant amount of energy deposited in excitations below the band gap, which are more immune to scattering. Alternatively, it can be read as an effective channeling depth of He in the material---channeling is absent from SRIM's amorphous-target model---changing the shape of the energy deposition profile; crystal orientations were not controlled or avoided during imaging, so channeling contributions are possible. The model and the data cannot distinguish between these mechanisms, and further work is ongoing to clarify this.

The back collection penalty is also reasonable: the asymmetric model returns $C_b = \qty{0.65(8)}{} < 1$, as expected since the geometry collects fewer SEs from the back, whereas the symmetric model requires a non-physical $C_b = \qty{3.0(2)}{}$; $C_b$ is prior-sensitive, tracking the assumed entrance scale (0.49--0.81 over the 1.5--2.5~nm sweep). The preference for the asymmetric model is conditional on these physical priors, not on fit quality alone: the free-$\lambda$ variant fits equally well statistically, but only with a shared $\approx$8~nm escape depth---four times the physical value---and a threefold back-collection advantage; the exit scale is deliberately the one left free, because exit-side transport at these velocities is the unknown process under study (Sec.~SVIII). In short: with local SE generation, an accurate SRIM profile, and shallow entrance escape, the data require asymmetric transport toward the exit surface.

Finally, both models converge to a similar stretch, $s = 0.70$ (symmetric) and $0.63$ (asymmetric), indicating a significant systematic difference between the EELS and SRIM depth scales; its size exceeds the 15--20\% EELS scale uncertainty alone (Sec.~SVIII). Part of the remainder may lie in SRIM itself, which obtains compound stopping by adding the stopping powers of the constituent elements (Bragg's rule), an approximation documented to degrade at low ion velocities and for non-metallic compounds.\cite{Bruckner2018}

The strong, reproducible thickness dependence suggests transmission-mode HIM as a rapid, low-fluence thickness measurement technique for suspended flakes, complementing EELS. Since the intensity rises then falls, a single value below the maximum is consistent with two thicknesses. Because the peak position tracks the ion range, which increases with beam energy, imaging at two accelerating voltages shifts the peak, which could make the pair of intensities single-valued (Sec.~SIX).

The correlation can also be inverted to probe transport: with thickness known from EELS, the modeled exit-side component reflects ion slowing, scattering, and near-surface SE transport.

In summary, the collected SE intensity of suspended \ce{MoS2} rises to a maximum near 40--55~nm at 30~keV and then decreases toward a weak top-surface signal, with an apparent back-to-front signal ratio of 4.7 at the maximum. Blocking and tape-converter tests favor the flake bottom surface, $\mathrm{SE1_{bottom}}$, as the leading source, and the energetic $\delta$-electron mechanism of fast-ion foils cannot operate here. The asymmetric SE-escape model applied to SRIM ionization profiles reproduces the peaked response ($\chi^{2}/\nu = 1.35$) with an effective exit-side escape depth of \qty{9.6(1.3)}{\nano\meter}, compared with the fixed 2~nm entrance-side scale, whereas the symmetric model does not; the added length may reflect forward-carried low-energy excitation or deposition physics beyond SRIM. The peaked correlation provides a low-fluence thickness screen and a route to testing ion--solid interaction models in thin, anisotropic materials.

\section*{SUPPLEMENTARY MATERIAL}
See the supplementary material for detector settings, fluence stability, and charging analysis; the copper-tape blocking, enhancement, and 45$^{\circ}$ converter geometries with the tape interception estimate and difference analysis; the EELS mean-free-path comparison; ROI counts and the Bayesian intensity model; the cross-geometry normalization of the blocked data; the chamber-geometry classification; the model definitions, the free-escape-depth variant, and fitting details; and the single-image thickness-precision estimate.

\begin{acknowledgments}
We are grateful for funding support from the Natural Sciences and Engineering Research Council of Canada (NSERC) and the 4D LABS core facility at SFU. DL is grateful for support from an NSERC Canada Graduate Research Scholarship—Doctoral (CGRS D).
\end{acknowledgments}

\section*{AUTHOR DECLARATIONS}
\subsection*{Conflict of Interest}
The authors have no conflicts to disclose.
\subsection*{Author Contributions}
\textbf{Cyan Kim}: Conceptualization (equal); Data curation (lead); Formal analysis (lead); Investigation (lead); Methodology (equal); Software (lead); Validation (equal); Visualization (lead); Writing -- original draft (lead); Writing -- review \& editing (equal). \textbf{David Lister}: Conceptualization (equal); Data curation (supporting); Formal analysis (supporting); Investigation (supporting); Methodology (equal); Software (supporting); Validation (equal); Visualization (supporting); Writing -- original draft (supporting); Writing -- review \& editing (equal). \textbf{Philip Jackle}: Investigation (supporting); Resources (supporting); Writing -- review \& editing (supporting). \textbf{Karen L. Kavanagh}: Conceptualization (equal); Funding acquisition (lead); Methodology (equal); Project administration (lead); Resources (lead); Supervision (lead); Writing -- original draft (supporting); Writing -- review \& editing (equal).

\section*{DATA AVAILABILITY}
The data and analysis code that support the findings of this study are openly available in the repository at \url{https://doi.org/10.5281/zenodo.21911088}.\cite{Kim2026dataset}
\bibliography{references}

\end{document}